# Improving the particle identification of radioactive isotope beams at the RIBLL2 separator

Fang Fang [1,2], Shuwen Tang [1,2], Shitao Wang [1,2], Xueheng Zhang [1,2,#],
Zhiyu Sun [1,2], Yuhong Yu [1,2], Duo Yan [1,2], Shuya Jin [1,2], Yixuan Zhao [1,2],
Shaobo Ma [1,2], Yongjie Zhang [1,2]

([1] Institute of Modern Physics, Chinese Academy of Science, Lanzhou 730000, China;
[2] University of Chinese Academy of Sciences, Beijing 100049, China)
[#] E-mail: zhxh@impcas.ac.cn

**Abstract:** To improve the ability of particle identification of the RIBLL2 separator at the HIRFL-CSR complex, a new high-performance detector for measuring fragment starting time and position at the F1 dispersive plane has been constructed and installed, and a method for achieving precise $B\rho$ determination has been developed using the experimentally derived ion-optical transfer matrix elements from the measured position and ToF information. Using the high-performance detectors and the precise $B\rho$ determination method, the fragments produced by the fragmentation of $^{78}$Kr at 300 MeV/nucleon were identified clearly at the RIBLL2-ETF under full momentum acceptance. The atomic number Z resolution of $\sigma_Z \sim 0.19$ and the mass-to-charge ratio A/Q resolution of $\sigma_{A/Q} \sim 5.8 \times 10^{-3}$ were obtained for the $^{75}$As$^{33+}$ fragment. This great improvement will increase the collection efficiency of exotic nuclei, extend the range of nuclei of interest from the A<40 mass region up to the A~80 mass region, and promote the development of radioactive nuclear beam experiments at the RIBLL2 separator.

## 1. Introduction

In-flight production of radioactive isotope (RI) beams by the projectile fragmentation method was pioneered in the 1980s at LBNL [1] and GANIL [2]. Since then, research on exotic nuclei using RI beams was advanced with the construction of in-flight fragment separators worldwide, including RIPS/BigRIPS at RIKEN [3, 4], A1200/A1900 at NSCL [5,6], FRS at GSI [7], and RIBLL at IMP [8]. With these facilities, the region of accessible exotic nuclei has been expanded significantly.

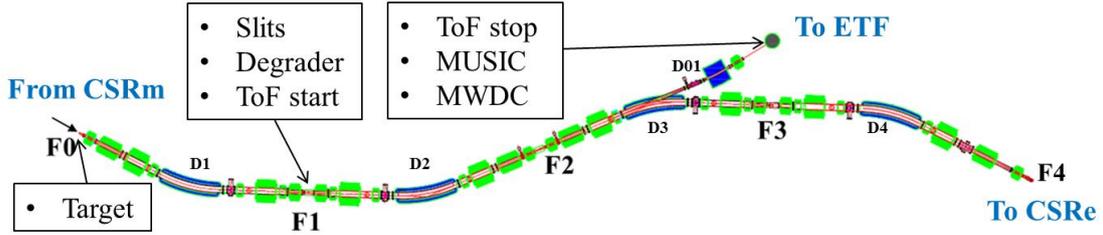

Figure 1 Schematic layout of the RIBLL2 separator.

To further expand research on exotic nuclei using RI-beams, a new in-flight fragment separator named the second Radioactive Ion Beam Line in Lanzhou (RIBLL2) has been constructed connecting the main cooler storage ring (CSRm) and the experimental storage ring (CSRe) at the Cooler-Storage Ring of the Heavy Ion Research Facility in Lanzhou (HIRFL-CSR) [9]. The schematic layout of the RIBLL2 is shown in Figure 1. The RIBLL2 is designed to be a double achromatic anti-symmetry spectrometer with a total length of 55 m. It has a maximum magnetic-rigidity of 10.64 Tm. Both the first half (F0-F2) and the second half (F2-F4) of the RIBLL2 are a mirror-symmetric system and can realize a point-to-point image, respectively. A maximum dispersion of 11.69 mm/% can be reached at the F1 and F3 dispersive planes. The resolving power of the separator is 1200 at the momentum deviation of $\Delta P/P \sim \pm 2\%$ and the divergence of $\pm 25$ mrad for a beam size at half width of 1mm.

The primary beams with several hundred MeV/nucleon energies extracted from CSRm will bombard the target at the F0. The RI-beams produced by projectile fragmentation will be collected and transported to different experimental terminals for physics researches. The RIBLL2 has been utilized to deliver RI-beams into the CSRe for mass measurements [10] and in-ring reaction measurement [11] using the fast extraction beams from the CSRm. Moreover, some cross sections for nucleon knockout reaction of exotic nuclei with a mass number up to 20 at energies ranging from 200 to 400 MeV/nucleon have been measured for probing the nuclear structure in combination with the first half of RIBLL2 and the external target facility (ETF) [12-14].

The particle identifications in these cross section measurement experiments were achieved with the $\Delta E$-ToF-$B\rho$ method in which the energy lose ($\Delta E$), time of flight (ToF), and magnetic rigidity ($B\rho$) are measured and used to determine the atomic number Z and the mass-to-charge ratio A/Q of the fragments. The layout of experimental setups for particle identification is schematically shown in Figure 1. A multiple sampling ionization chambers (MUSIC) [15] placed at the ETF was used to provide the $\Delta E$ information. The ToF was measured by the detectors installed at the F1 and ETF with a flight path of 26 m. The $B\rho$ value can be determined from the position measurements at the F1 and ETF foci.

The ToF start detector installed at the F1 consists of a photomultiplier tube (PMT) attached to a thin plastic scintillator with a dimension of $100 \times 100 \times 3$ mm$^3$ [16]. The large size is required to cover the large beam spot at the dispersive plane. For installation in tight spaces, the ToF start detector can only be read out using a single PMT from a single end. This makes the detector time resolution affected by the hitting positions of fragments. Also, there is no position detector in the F1 dispersive plane for the precise Bρ determination of fragments. Taking advantage of the current experimental setups, we got a poor mass resolution even for the light fragments produced by the fragmentation of $^{16}$O at 360 MeV/nucleon under full momentum acceptance conditions [12]. By reducing the momentum acceptance with a pair of horizontal slits at the F1, the fragments with a mass number up to 40 were identified at the RIBLL2-ETF [17]. This has reached the limit of particle identification with the current detectors at the RIBLL2-ETF in the case of guaranteeing sufficient fragment yields for physics studies.

The mass resolution $\sigma_A/A$ of fragments can be simply determined by,

$$\left(\frac{\sigma_A}{A}\right)^2 = \left(\frac{\sigma_{B\rho}}{B\rho}\right)^2 + \left(\frac{\gamma^2 \sigma_{TOF}}{TOF}\right)^2 \tag{1}$$

where $\sigma_M$ is the standard deviation of measured values of "M", and γ represents the relativistic Lorentz factor. To improve the mass resolution, the common method is to improve the Bρ accuracy and the time resolution of the ToF system. With the dispersion value of the RIBLL2, the $\sigma_{B\rho}/B\rho$ value can be assessed at ~0.0017 assume the position information with a 2 mm resolution can be obtained at the dispersive plane. If we plan to get a mass deviation of 0.25 for the fragment with a mass number of 80 and with a kinetic energy of 300 MeV/nucleon, the time resolution of the ToF system should be better than 198.4 ps according to Eq.(1). To satisfy the requirements of particle identification, a new high-performance detector for measuring fragment starting time and position at the F1 dispersive plane has been proposed for the RIBLL2-ETF separator.

In this paper, the configuration and operation of the new high-performance detector will be described together with the tests of this detector with the beams. The details of the particle identification scheme in the RIBLL2-ETF separator will also be presented.

## 2. Experiments and results

### 2.1 Configuration of the new ToF start detector

The new high-performance detector for measuring fragment starting time and position at the dispersive plane has been designed and constructed. Its schematic layout is shown in Figure 1(a). It is composed of a 100×100×1 mm$^3$ EJ 232 scintillator sheet and a plastic scintillator strip array with a sensitive area of 100×100 mm$^2$.

The scintillator sheet is read out on four sides, each by 10 silicon photomultipliers (SiPMs) from Hamamatsu Photonics (MPPC S13360-6075PE). The 10 SiPMs are mounted on a PCB and connected in series, which is good for getting a better time resolution because detected light yields increase and smaller capacitance makes a rise time faster [18]. The start time of fragments can be defined as a quarter of the sum of times measured by the SiPMs attached to the four sides of the scintillator sheet. This can effectively eliminate the effect of the fragment hitting positions on the time resolution.

The plastic scintillator strip array is placed downstream of the scintillator sheet. It has 50

detector modules, and each module consists of an EJ212 scintillator strip with a dimension of $120\times2\times0.5$ mm$^3$ and two SiPMs (MPPC S13360-3050PE) coupled to both ends for readout. The fragment position information at the F1 with a resolution of 2 mm in the horizontal direction can be obtained from this scintillator strip array.

All the SiPMs output signals from the scintillator sheet and the scintillator strip array is fed into a specific high-resolution time measurement module. It is a 6U PXI module and can measure the time information with a 25 ps time resolution for 16 SiPM signals simultaneously [19].

The new ToF start detector has been installed in the F1 dispersive plane at the RIBLL2. It is inserted into a vacuum box shown in Figure 2(b). This vaccum box has thin windows in both sides, and each window has an area of $100\times100$ mm$^2$ covered by a 0.1mm thick stainless steel foil. When inserted, the center of the new start detector will overlap with the center of two windows. The mobility of the detector is accomplished by connecting the vaccum box to the F1 chamber using a long bellows seal. By compressing the bellows, the detector can be moved within a range of 200 mm, which assures the detector is online or offline.

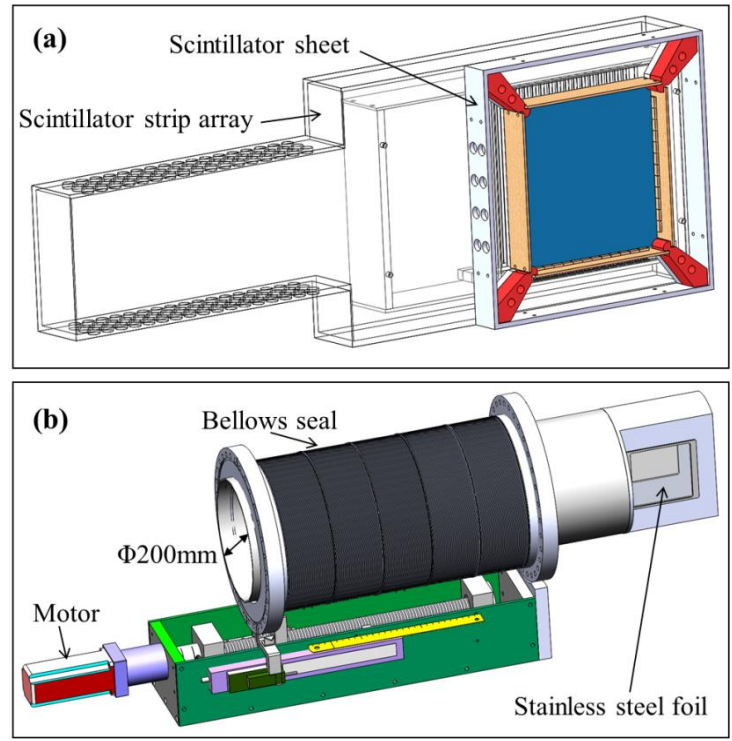

Figure 2 Schematic layouts of the new ToF start detector (a) and the vaccum box (b).

## 2.2 Projectile fragmentation experiment of $^{78}$Kr

The new ToF start detector was used for the time and position measurement at the F1 dispersive plane in the projectile fragmentation experiment of $^{78}$Kr at 300 MeV/nucleon. In this experiment, the primary beam was extracted from the CSRm and bombarded onto a 1.84 mm thick Al$_2$O$_3$ target at the F0 of the RIBLL2. The magnetic rigidity of the RIBLL2 separator was set to 5.44 Tm. At the F1 the degrader was not used, and the horizontal slits were fully opened, allowing more RI-beams to be delivered to the ETF by the first half of the RIBLL2.

At the ETF a 50×50×1 mm³ EJ200 scintillator sheet coupled to two H6410 photomultipliers from Hamamatsu Photonics at both ends was used as the ToF stop detector. Its signals were processed by the same electronics as the ones of the ToF start detector. In addition, a multi-wire drift chamber (MWDC) with an active area of 60×60 mm² was installed for fragment position measurements downstream of the ToF stop detector. It consists of 4×12 sense wires aligned in 4 layers. Each sense wire is surrounded by 2 field wires and some cathode wires, and these wires form a drift cell with a full width of 5 mm. The sense wires in the first two layers are sensitive to the horizontal direction. The wire positions for the second layers are shifted by a half cell width to solve the left-right ambiguities. The configuration of the sense wires in the last two layers is similar to the ones in the first two layers, but they are sensitive to the vertical direction. The induced signals from sense wires are integrated and discriminated by the SFE16 chips and then are digitized by the HPTDC chips [20].

**2.3 Experimental results**

In the experiment, the $^{78}$Kr beam passing through the target was utilized to test the performance of the new ToF start detector. Figure 3(a) shows the measured ToF spectrum, where the ToF is calculated by the time difference between the scintillator sheet detector at the F1 and the stop detector at the ETF. A room-mean-square (RMS) ToF resolution of ~221.9 ps was obtained. The main contributions to this ToF resolution include time resolution of time detectors, energy straggling of the beam in the target, and variance in the flight path length. According to ion-optical transformation, both the energy straggling and the flight path length is related to the fragment positions at the F1 dispersive plane. A correlation between the ToF and the position at the F1 is plotted in Figure 3(b), and a dependence of ToF on positions is observed. To correct the effect of position on ToF, we introduce the following formula,

$$ToF = ToF^{raw} + cx_{F1} \qquad (2)$$

Here ToF and ToF$^{raw}$ are the measured ToF and corrected ToF, respectively. The c is the correction coefficient. It can be obtained by linearly fitting Figure 3(b). Using this formula the ToF spectrum was corrected and shown in Figure 3(c). An obvious improvement in the time resolution from ~221.9 ps to ~186.7 ps is obtained after this correction. It is slightly better than the calculated value of Eq.(1).

Assuming that the intrinsic time resolution of the readout of both ends of a detector are the same, then for a detector illuminated by a small beam spot the width of the time difference between the times measured from both ends can indicate $\sqrt{2}$ times of the timing resolution of the detector. Figure 3(d) shows the time difference spectra between the times measured from the left and right sides of the ToF start and stop detectors in the $^{78}$Kr experiment. The horizontal beam spot size on the ToF start and stop detectors is limited to be 10 mm using the position information measured with the plastic scintillator strip array and MWDC detectors, respectively. According to the fitting results, the time resolution of the ToF start and stop detectors is evaluated to be ~58.7 ps and ~140.5 ps, respectively. The worse time resolution of the ToF stop detector is caused mainly by poor H6410 photomultiplier performance, which has a slow rise time (~2.7 ns) and large transit time spread (~1.1 ns). By using the fast time photomultiplier likes R2083 from Hamamatsu Photonics or SiPMs connected in series likes readout in the ToF start detector in place of H6410 photomultiplier, the time resolution of the ToF stop detector can be expected to be improved.

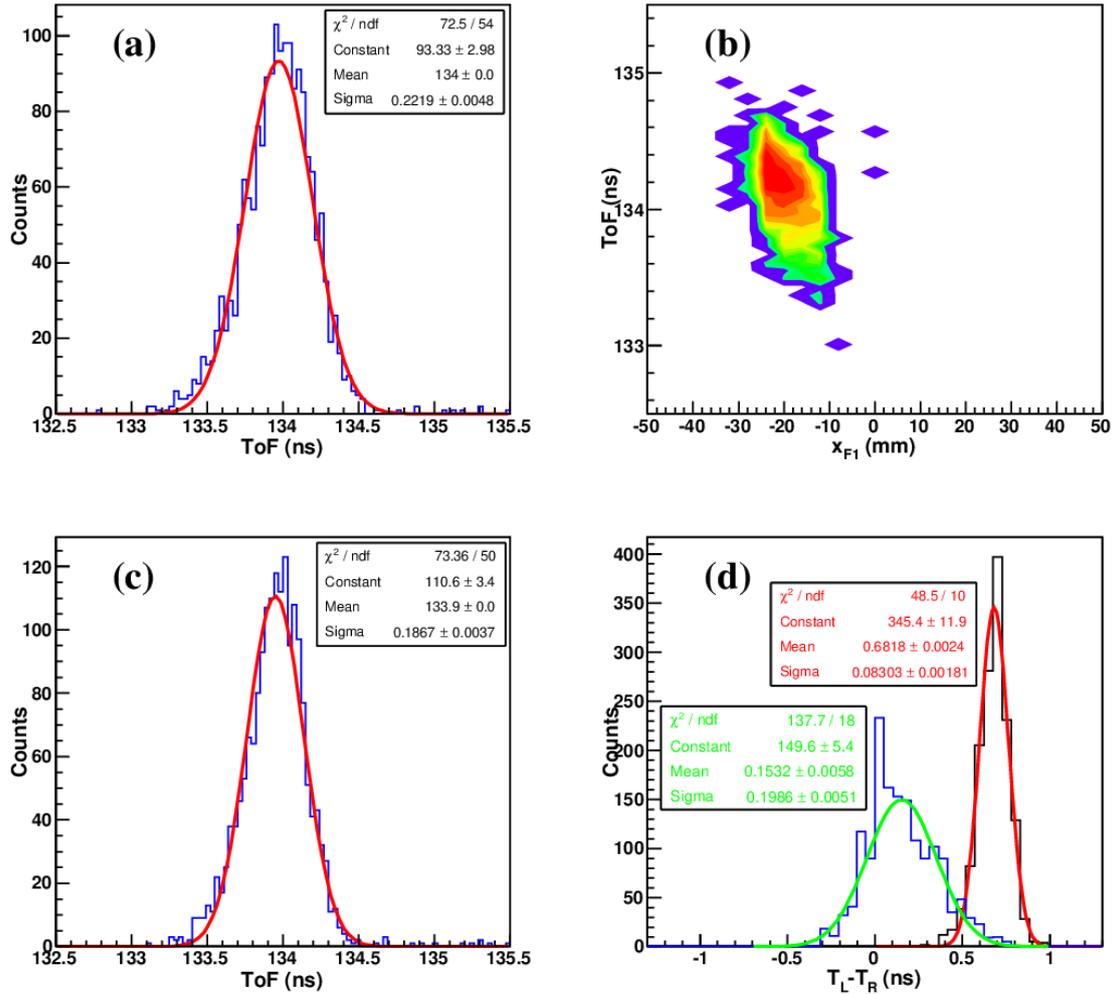

Figure 3 Performance of the new ToF start detector tested with the $^{78}$Kr beam derived from the passage of the primary beam through the Al$_2$O$_3$ target, (a) ToF spectrum obtained from the time difference between the ToF start and stop detectors, (b) Correlation between the ToF and the position at the F1 focal plane, (c) The ToF spectrum corrected with the positions, (d) Time differences between the left and right ends of the ToF start and stop detectors. All the lines in the figures denote the fitting results with the Gaussian function.

According to the coincidence measurements of ΔE, ToF, and Bρ, the atomic number Z and the mass-to-charge ratio A/Q of the fragments produced in the reaction $^{78}$Kr+Al$_2$O$_3$(1.84 mm) at 300 MeV/nucleon were determined and shown in Figure 4(a). In the calculation of A/Q, the fragment Bρ values were fixed at 5.44 Tm. From the results, it is obvious that all the elements can be clearly separated. The fragment charge spectrum obtained from the projection on the Z axis is shown in Figure 4(b). The charge resolution is approximately ~0.25 charge units (RMS) for the lower Z elements, which have small energy loss in the MUSIC. With the increase of the charge number, the charge resolution becomes slightly better and a good charge resolution of ~0.19 charge units (RMS) was achieved for the As element. Compared to the charge resolution, however, a poor fragment mass resolution was obtained even for the light fragments, and the isotopes from Ca to Se can not be identified unambiguously. This is mainly because of the large Bρ spread resulted from the fully opened slits at the F1. In this case, a precise Bρ measurement should be used instead of the constant value.

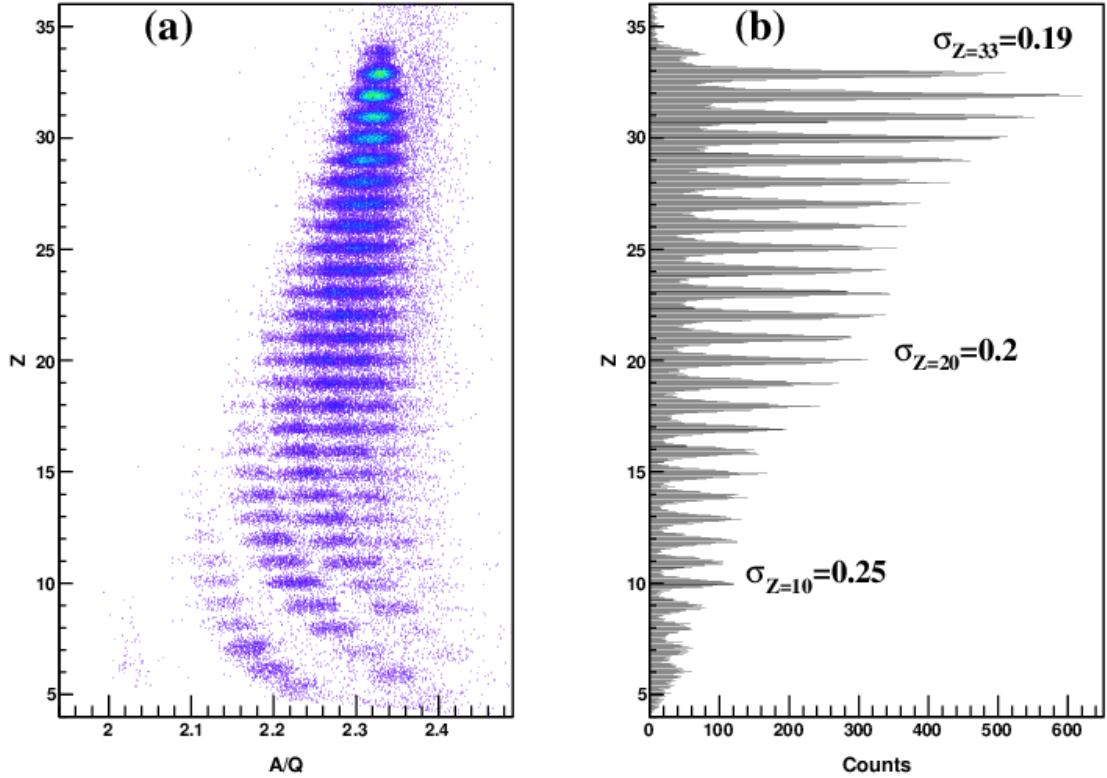

Figure 4 (a) Z versus A/Q particle identification plots for fragments produced in the reaction $^{78}$Kr+Al$_2$O$_3$(1.84mm) at 300 MeV/nucleon. In the calculation, the Bρ values were fixed at 5.44 Tm. (b) the fragment charge spectrum obtained from the projection on the Z axis.

In the case of the horizontal plane, the ion-optical transformation is given by:

$$x_1 = (x|x)x_0 + (x|a)a_0 + (x|\delta)\frac{B\rho - B\rho_0}{B\rho_0}. \qquad (3)$$

Here $x_1$ is the horizontal position at the image position. The $x_0$ and $a_0$ represent the horizontal position and angle at the object position. The first-order matrix elements $(x|x)$, $(x|a)$, and $(x|\delta)$ denotes the image magnification, the angular dependence, and the momentum dispersion, respectively. The $B\rho_0$ is the central magnetic rigidity. Using this relationship, the precise Bρ determination could be realized with the measured coordinate information at the object and image positions.

Using Eq.(3), the Bρ values have been determined with the designed matrix elements. For fragments transported from the F0 to the F1, the design values of (x|x), (x|a), and (x|δ) are 0.54, 0, and 11.69 mm/%, respectively. Assuming negligible spot size at the F0 object position, the Bρ values can be calculated simply from the following equation,

$$B\rho = B\rho_0 \left(1 + \frac{x_{F1}}{(x|\delta)}\right). \quad (4)$$

Here $x_{F1}$ represents the positions in the horizontal plane at the F1. Using the designed dispersive value of 11.69 mm/% and the horizontal positions measured with the new ToF start detector at the F1, the Bρ values of the fragments were determined and the A/Q values were recalculated and shown in Figure 5(a). All the isotopes from Be to Se can be distinguished obviously compared to Figure 4(a). But for the fragments in the region of Ca to Se, the mass resolution is still poor. That is because the designed (x|δ) value was used in the calculations. The difference between the designed and experimental momentum-dispersion values results in a non-negligible deterioration of the Bρ and mass resolutions.

According to the measured various correlations between the F1 and ETF foci, the experimental image magnification and momentum dispersion can be derived using Eq.(3). For example, in the experiment, the angular dependence (x|a) equals zero if the focusing is realized. The gradient of the correlation between the $x_{F1}$ and $x_{ETF}$ gives the (x|x) element if the events with (Bρ- Bρ$_0$)/Bρ$_0$≈0 are selected, and the correlation between the (Bρ- Bρ$_0$)/Bρ$_0$ and $x_{ETF}$ allow us to determine the (x|δ) element if the events with $x_{F1}$≈0 are selected.

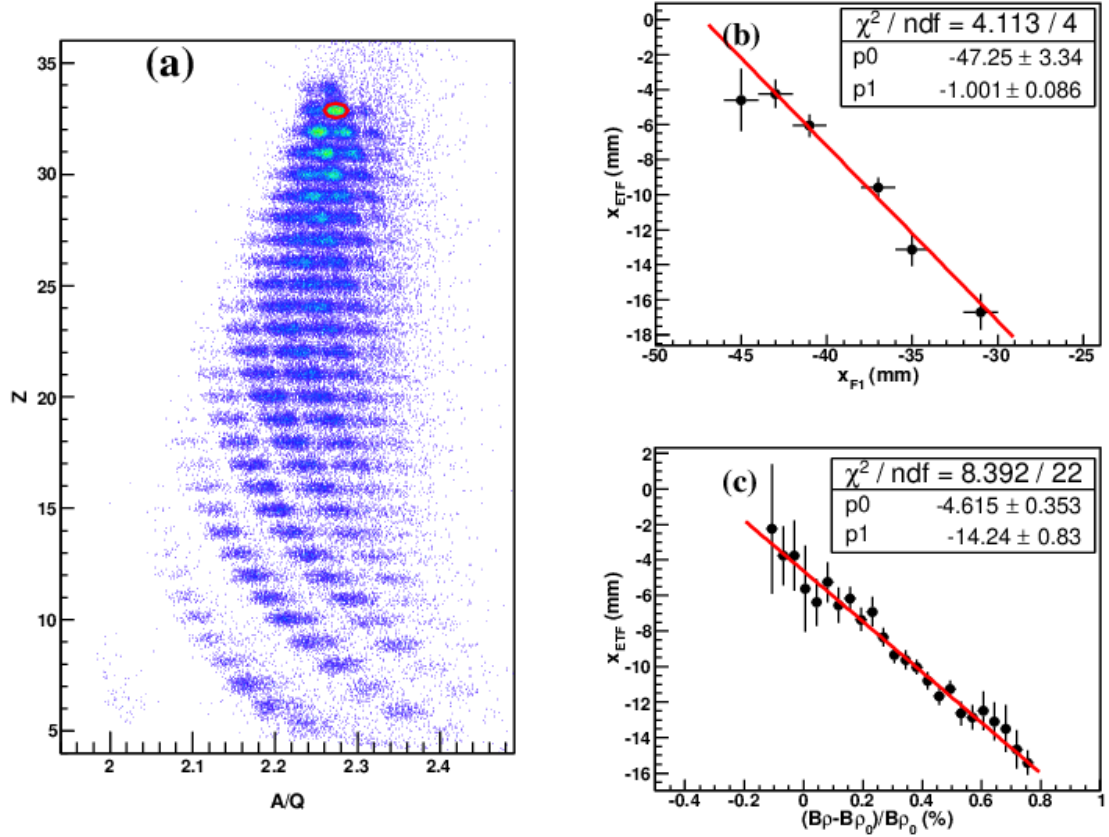

Figure 5 (a) Same as in Figure 4(a), but the Bρ values were calculated by Eq.(4) with the designed momentum dispersion. (b) Plot of the correlation between $x_{ETF}$ and $x_{F1}$ for (ToF- $ToF_0$)/$ToF_0$<0.02% with the selected isotope $^{75}As^{33+}$. The solid line shows the fitted line using a linear function. (c) Same as in Figure 5(b), but a plot of the correlation between $x_{ETF}$ and (Bρ- $Bρ_0$)/$Bρ_0$ for $x_{F1}$<2 mm.

To further improve the Bρ resolution, the derivation of the first-order matrix elements has been made in this $^{78}Kr$ fragmentation experiment with the selected isotope $^{75}As^{33+}$ which is circled with a solid line in Figure 5(a). The (Bρ- $Bρ_0$)/$Bρ_0$ values were obtained from the ToF measurement. The correlation between $x_{ETF}$ and $x_{F1}$ for (ToF- $ToF_0$)/$ToF_0$<0.02% and between $x_{ETF}$ and (Bρ- $Bρ_0$)/$Bρ_0$ for $x_{F1}$<2 mm with the selected isotope is plotted in Figure 5(b) and (c), respectively. The experimental image magnification of -1.0±0.09 and momentum dispersion of 14.24±0.83 mm/% are obtained with these correlations. The increase in momentum dispersion is mainly caused by the additional 9° dipole magnet near the ETF focus labeled as D01 in Figure 1. Using the experimentally derived matrix elements, the Bρ values of the fragments were recalculated and the A/Q values were again determined and shown in Figure 6(a). It is obvious that there is a significant improvement compared to Figure 5(a) and the fragments up to Se element can be identified unambiguously.

The A/Q spectra of Ne, Ca, and As isotopes are shown in Figure 6(b), (c), and (d), respectively. The fitting results display that the mass resolution gets better with the increase of fragment mass number. That is because in the data processing we focused mainly on heavy isotopes that are difficult to the identification. The $^{75}As$ isotope, which has higher yields, was selected for the derivation of the experimental matrix elements. This may cause the light mass isotopes to have poor mass resolutions. However, the current mass resolution is enough for the light isotopes. The absolute RMS A/Q resolution for $^{75}As^{33+}$ is as high as $σ_{A/Q}$~5.8e-3, revealing the importance of the derivation of experimental matrix elements for the precise Bρ determination.

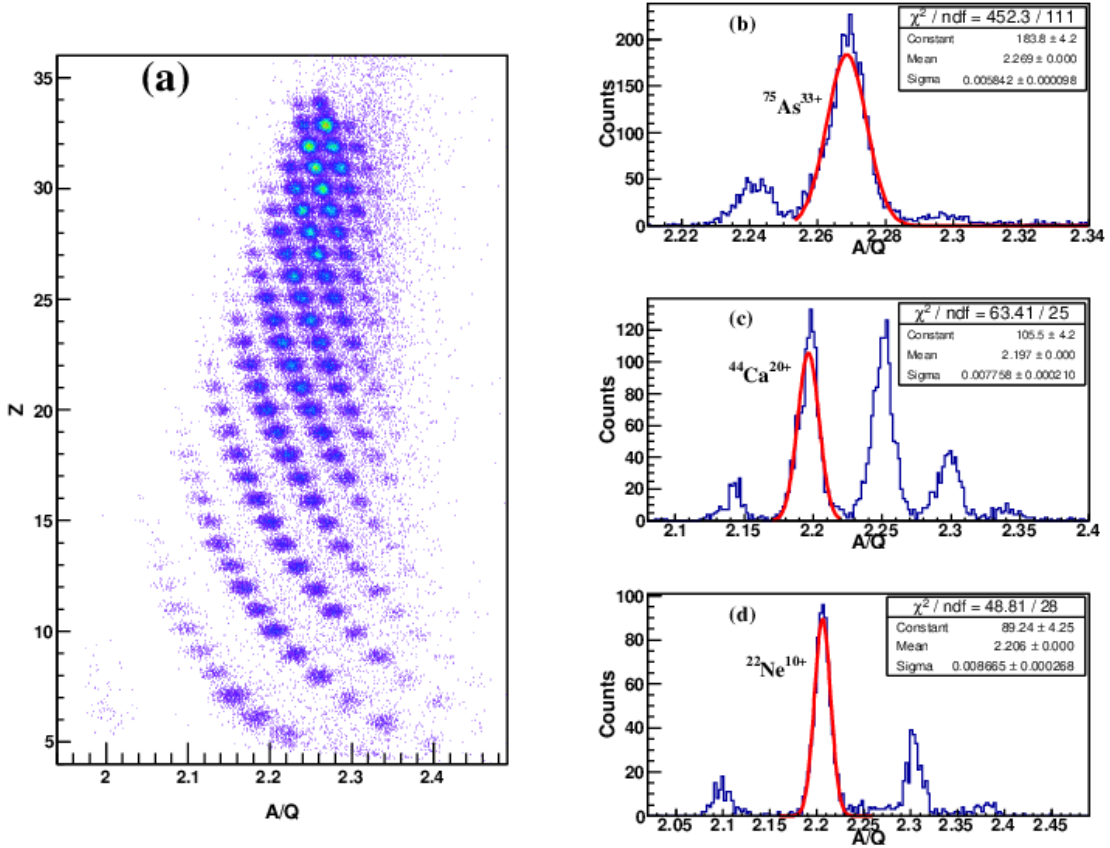

Figure 6 (a) Same as in Figure 4(a), but the Bρ values were determined by Eq.(3) with the experimental matrix elements in the calculation. (b), (c), and (d) show the A/Q spectra of Ne, Ca, and As isotopes, respectively.

In data processing, the effect of the angular dependence (x|a) in Eq.(3) was ignored. That is because only the position information can be used without angles. If the incident angles at the F1 and ETF foci can be measured in the future, the (x|a) element can be derived from the experimental data likes the (x|x) and (x|δ) elements. This may further improve the accuracy of the Bρ determination and get better particle identification.

## 3. Summary

With the help of the high-performance detectors and the precise Bρ determination method, the fragments produced by the fragmentation of $^{78}$Kr at 300 MeV/nucleon were identified clearly at the RIBLL2-ETF under full momentum acceptance. The atomic number Z resolution of $\sigma_Z \sim 0.19$ and the mass-to-charge ratio A/Q resolution of $\sigma_{A/Q} \sim 5.8 \times 10^{-3}$ were obtained for the $^{75}$As$^{33+}$ fragment. This great improvement will be a benefit to increase the collection efficiency of exotic nuclei and to extend the range of nuclei of interest from the A<40 mass region up to the A~80 mass region. However, the current A/Q resolution is still not ideal for the identification of the heavier fragments or fragment charge states. Some further improvement plans have been proposed. For example, the time resolution of the ToF system will be further improved by upgrading the ToF stop detector with the same scheme as the ToF start detector. Meanwhile, some new types of detectors on high-performance and compactness, such as the electrostatic-lens position-sensitive TOF MCP detector [21], will be developed to replace the existing ones. In addition, the beam optics of the RIBLL2 will be further optimized to improve the beam resolving power. Combined with improving the position resolution of the position detector and measuring the angle information at the F1 and ETF foci, the accuracy of the Bρ determination can be expected to be further improved. With these improvements, a better A/Q resolution would be achieved. Also, it would provide valuable particle identification experiences for the High energy FRagment Separator (HFRS) [22] at the High Intensity heavy ion Accelerator Facility (HIAF) [23].